 \documentclass[final,1p,times]{elsarticle}

\usepackage{amssymb}

\usepackage{amsmath,amsfonts}
\usepackage{algorithmic}
\usepackage{algorithm}
\usepackage{array}
\usepackage{textcomp}
\usepackage{stfloats}
\usepackage{url}
\usepackage{verbatim}
\usepackage{graphicx}
\usepackage{siunitx}

\usepackage{amsthm}

\usepackage{mathtools}
\usepackage{epstopdf}
\usepackage{multirow}

\usepackage{tikz}
\usetikzlibrary{positioning}

\usepackage{appendix}

\newcommand{\changes}[1]{\textcolor{black}{#1}}

\journal{Ecological Informatics}

\begin{document}

\begin{frontmatter}



\title{Soft-Output Signal Detection for Cetacean Vocalizations Using Spectral Entropy, K-Means Clustering and the Continuous Wavelet Transform}


\author[inst1]{M.W. Rademan}

\affiliation[inst1]{organization={Department of Electrical and Electronic Engineering},
            addressline={University of Stellenbosch}, 
            city={Cape Town},
            postcode={7600}, 
            state={Western Cape},
            country={South Africa}}

\author[inst1]{D.J.J. Versfeld}
\author[inst1]{J.A. du Preez}

\begin{abstract}
Underwater passive acoustic monitoring systems record many hours of audio data for marine research, making fast and reliable non-causal signal detection paramount. Such detectors assist in reducing the amount of labor required for signal annotations, which often contain large portions devoid of signals.

Cetacean vocalization detection based on spectral entropy is investigated as a means of vocalization discovery. Previous techniques using spectral entropy mostly consider time-frequency enhancement of the entropy measure, and utilize the short time Fourier transform (STFT) as its time-frequency (TF) decomposition. Spectral entropy methods also requires the user to set a detection threshold manually, which call for knowledge of the produced entropy measures.

This paper considers median filtering as a simple, effective way to provide temporal stabilization to the entropy measure, and considers the continuous wavelet transform (CWT) as an alternative TF decomposition. K-means clustering is used to determine the threshold required to accurately separate the signal/no-signal entropy measures, resulting in a  one-dimensional, two-class classification problem. The class means are used to perform pseudo-probabilistic soft class assignment, which is a useful metric in algorithmic development. The effect of median filtering, signal-to-noise ratio and the chosen TF decomposition are investigated. 

The accuracy and specificity measures of the proposed detection technique are simulated using a pulsed frequency modulated sweep, corrupted by a sample of ocean noise. The results show that median filtering is particularly effective for low signal-to-noise ratios. Both the STFT and CWT prove to be effective TF analyses for signal detection purposes, each presenting with different advantages and drawbacks. The simulated results provide insight into configuring the proposed detector, which is compared to a conventional STFT-based spectral entropy detector using manually annotated humpback whale (\textit{Megaptera novaeangliae}) songs recorded in False Bay, South Africa, July 2021.

The proposed method shows a significant improvement in detection accuracy and specificity, while also providing a more interpretable detection threshold setting via soft class assignment, \changes{providing a detector for use in development of adaptive algorithms.}
\end{abstract}



\begin{highlights}

\item \changes{A modified spectral entropy detector is proposed.}

\item \changes{The continuous wavelet transform is proposed as an alternative time-frequency decomposition.}

\item \changes{Soft-detection is produced by applying the k-means algorithm.}

\item \changes{The technique is evaluated in simulation and with real annotated data.}

\item \changes{The proposed detector is compared to a traditional spectral entropy detector.}

\item \changes{The proposed detector is superior with respect to accuracy and specificity measures.}

\end{highlights}


\end{frontmatter}



\section{Introduction}

Underwater passive acoustic monitoring (PAM) systems consisting of hydrophones are often used to record marine activity \cite{hydrophone, internalsignaldetection, dolphindetection}. \changes{PAM is a very effective tool which may aid in ecological monitoring, preservation and research. However, PAM systems present their own of set challenges relating to the field of digital signal processing.} The recorded vocalizations are often from distant sources, which may result in poor signal-to-noise ratios. PAM systems record hundreds of hours of audio from multiple hydrophones, making it time-consuming to find and label vocalizations manually. Thus, reliable and computationally efficient signal detection is of great importance for data gathering purposes.

Many cetaceans, such as the humpback whale (\textit{Megaptera novaeangliae}) and southern right whale (\textit{Eubalaena australis}), produce vocalizations that are periodic in nature. These signals may be modelled as amplitude-frequency modulated (AMFM) sinusoids \cite{whalemodel}. Recordings of these vocalizations are useful for conservation studies, like determining migration patterns or population estimates. It has become critical to monitor the effect of industry on marine mammals, and to take measures for ecological preservation. Numerous detection algorithms and monitoring systems have been developed to aid in this venture \cite{glider, rightdetection}.

\changes{Recent techniques which detect and/or classify cetacean calls often make use of machine learning (ML) models, using deep learning \cite{MLcnn, MLalexnet} or hidden Markov models (HMM) \cite{hmm1, hmm2}. Due to the difficulty of obtaining large amounts of annotated data, neural network ML models often need to use transfer learning as in \cite{MLalexnet}. ML techniques often require additional feature extraction methods to reduce the dimensionality of the audio data, thus adding to computational complexity \cite{hmm3}. ML models cannot necessarily operate well outside the conditions of their training data. Training ML models may also be a cumbersome task \cite{unsupervised}, making simpler, unsupervised techniques which are more computationally efficient the preferred method of sifting through large amounts of data to identify points of interest. These points of interest allow biologists to ignore extensive portions of noise in the audio, thus greatly reducing manual work, before classifying the audio.}

\changes{An example of such an efficient detector which is prevalent in PAM audio analysis software packages (PAMGuard \cite{pamguard} and Raven Pro \cite{raven}), is the band-limited energy detector (BLED) \cite{internalsignaldetection}. This detector is simple to compute, making it a popular choice for finding points of interest in long audio recordings. BLED has also seen use as part of multi-stage classifier systems \cite{unsupervised}.}

\changes{Spectral entropy (SE) is another popular and simple method derived from Shannon's information entropy \cite{shannon} to perform narrow-band signal detection, and has demonstrated success in both speech and cetacean vocalization detection, consistently outperforming BLED \cite{entropyJASA, entropyOCEANS, entropysubband}}.  At its core, spectral entropy utilizes a time-frequency (TF) decomposition and treats each time-slice as a pseudo-probability density. SE will decrease in the presence of narrow-band frequency content. This method of narrow-band signal detection was first proposed by Powell and Percival in \cite{entropyorig}. 

In real data, the SE measure may fluctuate due to changing noise conditions, resulting in less accurate detection. Various methods for speech detection aim to stabilize/enhance the SE as in \cite{entropyOCEANS, entropysubband}, which utilize weighting factors in the SE calculation \changes{as well as temporal averaging. For signal detection, the researcher must determine the appropriate SE threshold manually, which again increases labor, and requires some understanding of how to set the detection threshold properly}. Furthermore, \cite{entropyOCEANS, entropysubband} only consider the short-time Fourier transform (STFT) as a TF decomposition. 

\changes{The continuous wavelet transform (CWT) has shown success in many applications, including feature extraction in ML models for electrocardiogram classification \cite{ecgcwt}, analysis of periodic patterns in lake water levels \cite{lakelevelcwt} and detecting pachyderm movements using seismic sensors \cite{seismiccwt}. The CWT has previously been used in structural engineering applications as a TF decomposition for SE \cite{cwtentropy1, cwtentropy2}, and may be considered as an alternative to the STFT for SE calculation.}

Erbe and King \cite{entropyJASA} used SE to detect  cetacean calls in the Canadian Arctic. They demonstrate its effectiveness compared to BLED, which SE outperforms on all accounts. Erbe and King used the simplest form of SE with the STFT, without any enhancement of the entropy measure, which will serve as a baseline comparison for this paper.

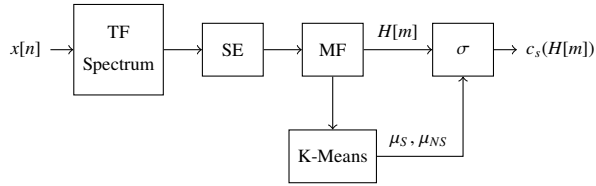
\begin{figure}[h!]
	\centering
	\begin{tikzpicture}
		
		\node (x) at (0,0) {\scriptsize $x[n]$};
		
		\node[draw,
		fill=white,
		minimum width=0.8cm,
		minimum height=1.2cm,
		align=center,
		right = 0.3cm of x
		] (TF) {\scriptsize TF\\ \scriptsize  Spectrum};
		
		\node[draw,
		fill=white,
		minimum width=0.8cm,
		minimum height=0.7cm,
		right= 0.5cm of TF
		] (SE) {\scriptsize SE};
		
		\node[draw,
		fill=white,
		minimum width=0.8cm,
		minimum height=0.7cm,
		right= 0.5cm of SE
		] (MF) {\scriptsize MF};
		
		\node[draw,
		fill=white,
		minimum width=0.8cm,
		minimum height=0.7cm,
		right= 0.9cm of MF
		] (Sigmoid) {\scriptsize $\sigma$};
		
		\node[draw,
		fill=white,
		minimum width=0.8cm,
		minimum height=0.7cm,
		below= 0.7cm of MF
		] (KM) {\scriptsize K-Means};
		
		\node (cs)[right=0.3cm of Sigmoid] {\scriptsize $c_s(H[m])$};	
		
		\draw[->] (x) -- (TF.west);
		\draw[->] (TF.east) -- (SE.west);
		\draw[->] (SE.east) -- (MF.west);
		\draw[->] (MF.east) -- (Sigmoid.west)node[above, midway]{\scriptsize $H[m]$};
		\draw[->] (Sigmoid.east) -- (cs.west);
		\draw[->] (MF.south) -- (KM.north);
		\draw[<-]  (Sigmoid.south) |- (KM.east) node[above, near end]{\scriptsize $\mu_S$, $\mu_{NS}$};
		
	\end{tikzpicture}
	\caption{Block diagram of the proposed soft-output detector.}\label{fig:systdiag}
\end{figure}

This paper compares the STFT and CWT as TF decompositions for the SE calculation, and proposes median filtering (MF) as a simple, yet effective, method of providing temporal stabilization to the SE measure. Additionally, a means of converting the SE to a pseudo-probability is proposed, using the k-means algorithm as a computationally simple way to distinguish between signal/noise SE segments. The pseudo-probability measure is more interpretable than SE, and is automatically scaled, making it robust against changes in ambient ocean conditions across multiple recorded files, which will typically require manually resetting the SE threshold.  Refer to figure \ref{fig:systdiag} for a description of the proposed solution.

\changes{To test the accuracy and specificity of our proposed solution compared to the baseline SE detector, this paper measures detector performance on a per-sample basis on the audio. Simulation is used to verify and design the proposed technique, and gain some insight on how SNR and hyper-parameters influence the detector in a controlled test environment. The proposed detector is also verified with manually annotated data, to demonstrate that it performs well in a practical setting. Per-sample accuracy is a necessary measure to test detectors, since software packages (PAMGuard; Raven Pro) allow the user to discard detections of short length. Thus, detecting only part of the signal is not suitable for all applications.}

We demonstrate empirically that the L1-normalized CWT is a superior TF-decomposition for similar practical applications for signal detection using SE. Median filtering is shown to be a simple and viable form of SE stabilization to improve signal detection accuracy and specificity. Additionally, we propose the K-means algorithm to provide appropriate scaling for the SE to be converted into a pseudo-probabilistic measure of signal presence, which may be used to develop adaptive algorithms \changes{and provide more intuitive threshold settings for non-experts. To the authors' knowledge, this paper is the first to introduce a soft-classification procedure for SE detectors through the k-means algorithm. Additionally, to the authors' knowledge, it is the first time the CWT is used in a SE detector for cetacean detection applications.}

\changes{Improving SE-based detectors are of great importance for performing narrow-band signal detection. SE detectors offer good accuracy in identifying periodic signals, while requiring no training, and offering some robustness against fluctuation noise profiles. Depending on the TF decomposition parameters used, SE allows for accurate time-localized detections, which yields more detailed analysis on vocalization statistics, such as call length, and distance between calls, and may assist in creating accurate datasets via data augmentation for supervised ML models.}

\section{False Bay Data}
The data used in this paper was recorded via a hydrophone at False Bay, on 1 July 2021 at midnight. Two humpback whale songs were extracted in the time frame between 00:00 and 00:30 hours. The hydrophones were anchored 6 kilometers off the coast. The hydrophone capsule recorded at 15 metres below sea level. The audio is sampled 32 kHz, which is resampled to 2 kHz. This is adequate for preserving the observed Humpback whale vocalizations, which consist of songs in the 150-1000 Hz band. The audio is normalized to have an average power of 1.

\begin{figure}[h!]
	\centering
	\includegraphics[width=0.6\textwidth]{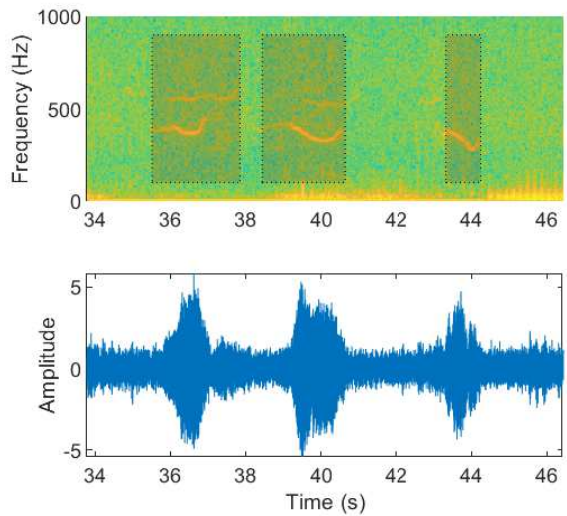}
	\caption{Audio snippet and spectrogram of a humpback song recorded in False Bay. The waveform shown is high-passed at 100 Hz, resampled and normalized. The signal annotations are shown on the spectrogram. The spectrogram is calculated before high-passing the signal.}
	\label{fig:fbdata1}
\end{figure}

Figure \ref{fig:fbdata1} shows a snippet of a recorded whale song. Many noise types can be observed. Low frequency anchor noise, caused by the swaying of the hydrophone in the water due to the anchoring system, contains the majority of signal power. This significantly corrupts all frequency content below 100 Hz. Thus, this noise is filtered out, since it does not influence the signals of interest, and is highly dependent on the anchoring system and the tides.

The remaining noise types are background noise, and impulsive noise, caused by snapping shrimp and fish. All other noise sources, such as waves, fish and boats are classified as background noise. 

Due to the dominating anchor noise, the signal-to-noise ratio (SNR) is only calculated for frequencies above 150 Hz to the Nyquist limit (1 kHz). The recorded whale calls are estimated to have an average SNR of approximately 0 dB, which is calculated in the time-frequency regions of signal presence, utilizing a power spectrogram. The instantaneous SNR is estimated to range from -20 to 30 dB.


The songs were manually annotated. The starting and ending time of a vocalization is recorded, which therefore only indicates signal presence. Overlapping calls are annotated as one vocalization. The two songs have a combined length of 369 seconds, containing 120 annotated calls. The signals compose 32\% of the recorded samples within the song segments. The remaining 68\% of samples are considered as noise.

\begin{figure}[h!]
	\centering
	\includegraphics[width=0.7\textwidth]{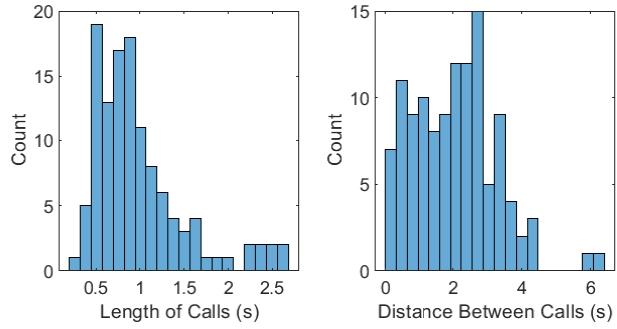}
	\caption{Histograms indicating the combined call length and distance between calls of the two humpback songs, for use in determining median filter window length limits.}
	\label{fig:fbhist}
\end{figure}

Figure \ref{fig:fbhist} show the combined statistics of the humpback songs. Call length and the distance between calls are calculated from the manual annotations. The distance is measured between the endpoint of a vocalization and the start of the next vocalization. These statistics are useful in determining median filter window lengths limits.

\section{Spectral Entropy}
\subsection{STFT}
Given a discrete-time signal $x[n]$, the windowed STFT may be expressed as

\begin{equation*}
	X[k, m] = \sum_{n=m}^{m + N - 1} w[m-n] x[n] e^{-j 2 \pi k n / N },
\end{equation*}

\noindent where $w$ is the windowing function, $N$ is the window size, $k \in \{0, 1, ..., N-1\}$ is the frequency index and $m$ is the time index. The total number of frequency bins may be artificially increased by padding each window with additional zeros.

The power spectrum $S[k, m]$ may be derived from $X[k, m]$ as

\begin{equation*}
	S[k,m] = \frac{1}{N}X[k,m] X^*[k,m],
\end{equation*}

\noindent which may then be converted to a pseudo-probability density $P[k,m]$, referred to as the spectral distribution:
\begin{equation}
\label{eqn:specdist}
	P[k,m] = \frac{S[k,m]}{\sum_{j=0}^{N-1} S[j,m]}.
\end{equation}

The spectral distribution is used to construct a time-varying entropy measure $H[m]$, based on Shannon's entropy:
\begin{equation}
\label{eqn:entropy}
	H[m] = - \sum_{k=0}^{N/2} P[k,m] \log (P[k,m]).
\end{equation}

It is well known that the uniform distribution is the maximum entropy distribution on a finite interval \cite{maxentropy}.  For spectral entropy, this corresponds to the spectral distribution of white noise. In a sense, it is a measure of certainty for discrete probability distributions: a flat spectral distribution will have high entropy, whereas the spectral impulse from a sinusoid will have low entropy. Note that the summation only occurs to $N/2$ (even $N$ assumed), due to the symmetry of the Fourier transform.

\begin{figure}[!t]
	\centering
	\includegraphics[width=0.6\textwidth]{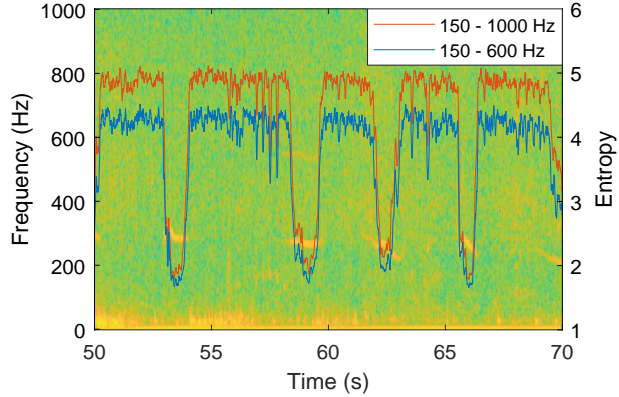}
	\caption{Spectral entropy (STFT) calculated for different frequency bin ranges.}
	\label{fig:entropy1}
\end{figure}

If a sinusoidal signal is known to be within a certain frequency range $f_0 \in (f_1, f_2)$, it may be beneficial to restrict the entropy measure between frequency indices:
\begin{equation*}
	H[m] = - \sum_{k=k_1}^{k_2} P[k,m] \ln (P[k,m]),
\end{equation*}

\noindent where $k_1 = \lfloor N f_1/f_s\rfloor$ and $k_2 = \lfloor N f_2/f_s\rfloor$, with $f_s$ representing the sample frequency. This will reduce the risk of signals within other bands contaminating the entropy measure. However, small frequency ranges are not advised, since a smaller frequency range will decrease the apparent ``flatness" of the measurement, hence lowering the entropy difference between signal/no-signal segments.


\subsection{CWT}
The CWT is a form of TF decomposition which falls under the theory of multi-resolution analysis (MRA). Wavelet decomposition is a topic of many facets and contain a great deal of mathematical depth. Mallat \cite{waveletbook} describes wavelet theory as an amalgamation of ideas from a variety of scientific fields, which were developed simultaneously.
 
From a signal processing point of view, the numerically calculated CWT may be interpreted as a set of finite impulse response (FIR) band-pass filters constructed to have a certain bandwidth and time support. This allows for more flexibility in the time-frequency resolution, depending on the chosen filter-bank parameters. The mother wavelet $\psi$ may be considered as the main impulse response (IR) of the filters, which is scaled in time such that its centre frequency coincides with a frequency of interest.

The analysing wavelet used in this paper is the generalized Morse wavelet, described by Olhede and Walden in \cite{morsewavelets, morsewavelets2}, which is defined by its Fourier transform:
\begin{equation*}
	\mathcal{F}\{\psi_{\beta, \gamma} (t) \} = \Psi_{\beta, \gamma}(\omega) = U(\omega) \alpha_{\beta, \gamma} \omega ^ \beta e^{-\omega \gamma},
\end{equation*}

\noindent with $\alpha_{\beta, \gamma} = (\frac{2e \gamma}{\beta})^{\beta/\gamma}$ as a normalizing constant, such that 
\begin{equation}
	\label{eqn:cwt:bpmag}
\Psi(\omega_{\beta,\gamma}) = 2,
\end{equation}

\noindent where $\omega_{\beta,\gamma} = (\frac{\beta}{\gamma})^{1/\gamma}$ is the peak frequency. The time-bandwidth product of the wavelet $P^2 = \beta \gamma$ determines the wavelet time support, which is proportional to $P = \sqrt{\beta\gamma}$. $U(\omega)$ represents the Heaviside step function:
\begin{equation*}
	U(\omega) = \begin{cases}
		1, & \omega > 0 \\
		0, & \omega \le 0
	\end{cases},
\end{equation*}

\noindent which removes all negative frequency components, making the Morse wavelet analytic (contains one-sided frequency information).

The frequency symmetry of the wavelet is controlled by $\gamma$. A symmetric Morse wavelet has $\gamma = 3$. $\beta$ further controls the time support of the wavelet, or equivalently, the filter bandwidth and IR length.

The CWT of a signal $x(t)$ is defined as:
\begin{equation}
	\label{eqn:cwt}
	X(s,b) = \frac{1}{\sqrt{|s|}} \int_{-\infty}^{\infty} x(t) \psi^*\left(\frac{t-b}{s}\right) dt,
\end{equation}

\noindent where $s$ is the scale parameter which expands/contracts the analyzing wavelet and $b$ is the time-shift parameter.  $\psi^*$ denotes the complex conjugate of $\psi$. Equation (\ref{eqn:cwt}) corresponds to band-pass-filtering $x(t)$ with the impulse response described by the scaled wavelet. The factor $|s|^{-1/2}$ provides L2 normalization, ensuring that energy is preserved with scaling.

In this paper, the MATLAB wavelet toolbox is used to perform CWT and frequency analysis. More information on how this toolbox performs CWT calculation may be found at \cite{cwtmatlab}. The number of band-pass filters is specified by the number of voices per octave, which result in a discrete set of scale parameters $\{s_1, s_2, ..., s_N\}$ which are related to one another by a factor $s_{n+1} / s_n = \sqrt[p]{2}$. This toolbox limits the ratio of the time/bandwidth parameters to $\beta_{max} = 40\gamma$. 

Given a starting frequency $f_1$, then
\begin{equation*}
	s_1 =  \frac{\omega_{\beta,\gamma}}{2\pi  f_1 } .
\end{equation*}

When calculating the CWT numerically, the MATLAB wavelet toolbox instead uses L1 normalization of the wavelets, thus preserving amplitude and not energy. The  scaling factor correction $|s|^{-\frac{1}{2}}$ in (\ref{eqn:cwt}) is therefore replaced by a factor of $|s|^{-1}$ in the MATLAB implementation. As such, all band-pass filters have a maximum gain of $2$ which compensates for the analytic nature of the Morse wavelet, since the combined positive and negative frequency components of a real sinusoid captures the amplitude.

The L1 normalized CWT is defined as
\begin{equation}
	X(s,b) = \frac{1}{|s|} \int_{-\infty}^{\infty} x(t) \psi^*\left(\frac{t-b}{s}\right) dt.
\end{equation}

\begin{figure}[t]
	\centering
	\includegraphics[width=0.7\textwidth]{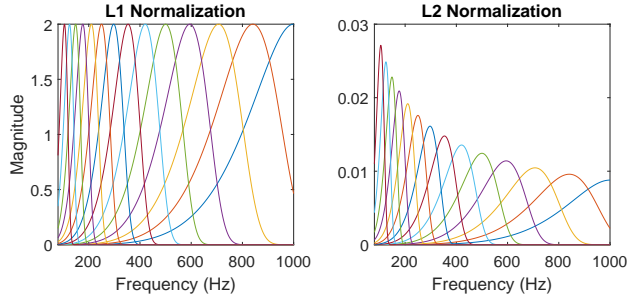}
	\caption{An example of a CWT filterbank, shown as the L1 and L2 normalized versions , with $\gamma = 10$, $\beta=50$ and 4 voices per octave.}
	\label{fig:cwt_fb_example}
\end{figure}

More information on the fast numerical calculation of the CWT may be found in \cite{fcwt}. Figure \ref{fig:cwt_fb_example} shows an example of the resulting MATLAB Morse wavelet filterbank (L1-normalized), and its L2-normalized version.

Given then the discretized/numerical calculation of the CWT with L1 normalization, $X[k,m]$, where $m$ denotes the discrete-time index of the transform, and $k$ denotes the index corresponding to $s$ which refers the centre frequencies ($f_c$) of the band-pass filters, a new pseudo-power spectrum may be utilized for entropy calculation. For a sinusoid with amplitude $A$ and a frequency corresponding to any $k_0$ of the band-pass filters' centre frequencies, the resulting amplitude of $|X[k_0,m]|$ will be $A$ from (\ref{eqn:cwt:bpmag}). Thus, the power of a sinusoid observed within a band is
\begin{equation}
	\label{eqn:cwtps}
	S[k,m] = \frac{1}{2} X[k,m] X^*[k,m]
\end{equation}

\noindent since the average power of a sinusoid with amplitude $A$ is $A^2/2$. For the CWT, (\ref{eqn:cwtps}) refers to the equivalent average power of a sinusoid at time index $m$ and frequency band $k$.

The bandwidth of the filter influences the observed power, especially in the presence of noise. However, the equivalent sinusoidal power may be more useful in most applications. When a sinusoid is present, the equivalent sinusoidal power does not reduce as frequency increases (as with L2 normalization). However, higher frequency wavelets will be more sensitive to noise, due to their increased bandwidth.

SE is calculated by substituting the CWT equivalent sinusoidal power from equation (\ref{eqn:cwtps}) into equations (\ref{eqn:specdist}) and (\ref{eqn:entropy}).

\section{Signal detection}
\subsection{K-means}
Many variants of the k-means algorithm were independently developed by researchers in the 1950's \cite{kmeansoverview}. The name was first coined by MacQueen in \cite{kmeans}. It has a rich history as a method of unsupervised classification, used in a wide set of applications. Its simplicity and interpretability make it a popular choice for many classification problems. \changes{In HMM classifiers, the k-means clustering algorithm is often used to initialize Gaussian mixture models (GMM) prior to data fitting \cite{hmm1, hmm2, hmm3}.} In this paper, k-means is used to determine the class means of the entropy measure given signal/no-signal conditions. This eliminates the need for a user-set threshold on the entropy and provides class means which may be used to construct a pseudo-probabilistic measure of signal presence. \changes{Refer to appendix \ref{app:kmeans} for details on the implementation for this application.}

\subsection{Soft Output}\label{sec:softoutput}

Given the class means $\mu_S$ (signal), $\mu_{NS}$ (no signal) determined by k-means clustering, the k-means decision boundary $\beta$ is the midpoint:
\begin{equation*}
	\beta = \mu_S + \frac{\mu_{NS} - \mu_S}{2}
\end{equation*}

In classical K-means, the assigned class $c_h(H[m])$ results in a hard output:
\begin{equation*}
	c_h(H[m]) = \begin{cases}
		1, & H[m] \leq \beta \\
		0, & H[m] > \beta
	\end{cases}
\end{equation*}

\noindent where a value of $1$ corresponds to signal presence and a value of $0$ representing no signal.

To provide a soft output, a function which smoothly transitions from one class to another in the interval $H[m] \in [\mu_S, \mu_{NS}]$ is required. Since an entropy measurement which is lower than $\mu_S$ is more probable to contain signal and, conversely, an entropy measurement which is greater than $\mu_{NS}$ is more probable to be noise, the transition function $\alpha(t)$ should satisfy the following criteria:
\begin{enumerate}
	\item  $\alpha(t)$ is monotonically decreasing.
	\item Approaches hard classification in the limit: $\lim\limits_{t \rightarrow \infty} \alpha(t) = 0$ and $\lim\limits_{t \rightarrow -\infty} \alpha(t) = 1$.
	\item Equal class probability at the boundary: $\alpha(t - \beta) |_{t=\beta} = \alpha(0) =  \frac{1}{2}$ .
	\item Class assignment symmetry: $\alpha(t) = 1 - \alpha(-t)$.
\end{enumerate}

The sigmoid function $\sigma(x)$ evaluated at $x = -t$ satisfies all the above criteria, defined as
\begin{equation*}
	\sigma(x) = \frac{1}{1 + e^{-x}}.
\end{equation*}

$\sigma(x)$ is proposed in the style of logistic regression \cite{logit}. A piecewise-linear function may also satisfy the above criteria.

\begin{figure}[!t]
	\centering
	\includegraphics[width=0.6\textwidth]{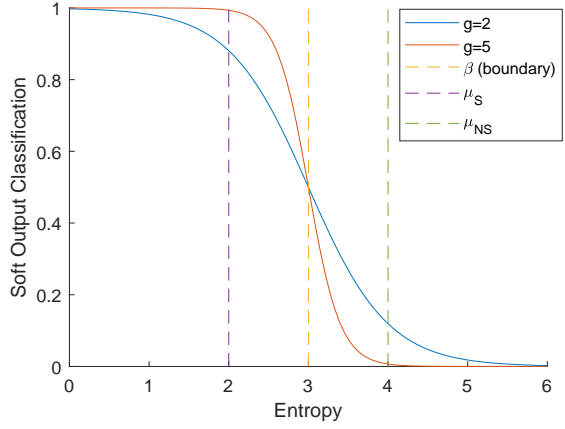}
	\caption{Example of the sigmoid function for soft output classification for various gains ($\mu_S=2$, $\mu_{NS}=4$).}
	\label{fig:sigmoid}
\end{figure}

The shape of the sigmoid function may be set by specifying the gain $g > 0$ as a hyper parameter. A higher gain will result in a steeper shape, thus attributing more certainty to entropy measurements closer to the class means, and vice versa for lower gains. It may also be beneficial to normalize the entropy measure such that $H[m] = \mu_S$ and $H[m] = \mu_{NS}$ map to $\pm1$. The normalization will ensure consistency in the behaviour of the hyper parameter $g$ over a range of class means. Figure \ref{fig:sigmoid} shows the effect of $g$ on the classification output.

The soft-output classification function $c_{s}(H[m])$ is therefore
\begin{equation*}
	c_{s}(H[m]) = \sigma\left(	-2g	\frac{H[m] - \beta}{\mu_{NS} - \mu_S}	\right).
\end{equation*}

To assign a specific pseudo-probability $c_{s}(H[m]) = p$ that corresponds to signal presence when $H[m] = \mu_S$, then the gain hyper parameter $g$ is calculated as
\begin{equation}
	\label{eqn:sigmoidg}
	g = \frac{1}{2} \frac{\mu_{NS} - \mu_S}{\mu_S - \beta} \ln\left(\frac{1}{p} - 1\right)
\end{equation}

For $g>0$, equation (\ref{eqn:sigmoidg}) requires $p > \frac{1}{2}$.

\subsection{Median Filtering}
Median filtering (MF) is often used to suppress ``salt-and-pepper" noise in image processing, and shown past success as a simplistic method of denoising images and signals \cite{mf1, mf2}.

Median filtering  may assist in suppressing the fluctuations of instantaneous spectral entropy measures prior to K-means, but preserve the ``step responses" present in the entropy measure when transitioning from signal to no-signal and vice versa. The entropy of background noise in practical data may not be as consistent as the entropy of GWN. It will therefore help suppress any outliers within the filtering window.

If the window is chosen large enough, it will suppress the entropy fluctuations of signals that are shorter than the half the window length. This may be useful if a restriction on signal length is imposed.

Given a discrete-time signal $x[n]$, and an window of odd length $M$, the median-filtered signal $y[n]$ is
\begin{equation*}
	y[n] = \text{med}(x[n - R], ..., x[0], ..., x[n + R] ),
\end{equation*}

\noindent where $R = \frac{M-1}{2}$ is the half-window length, and $\text{med}(x_1, ..., x_M)$ refers to the median of a sequence of $M$ numbers $x_1$ to $x_M$. Note that if $M=1$, then no median filtering is applied.

\begin{figure}[h!]
	\centering
	\includegraphics[width=0.6\textwidth]{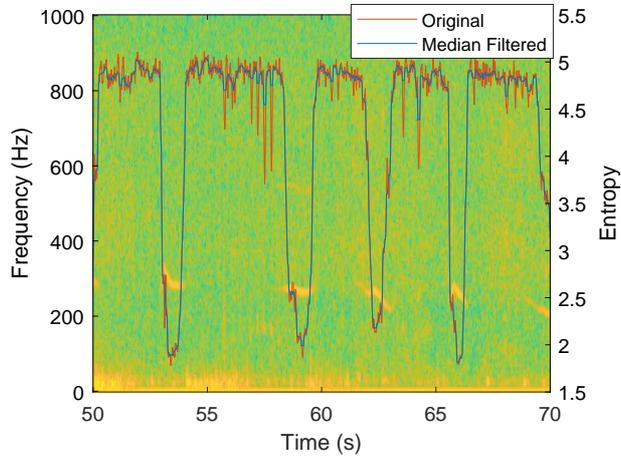}
	\caption{Median filtered entropy (STFT) with $M=355$.}
	\label{fig:entropymf}
\end{figure}

Figure \ref{fig:entropymf} shows an example of applying median filtering ($M=355$) on the STFT spectral entropy.

\begin{figure}[h!]
	\centering
	\includegraphics[width=0.6\textwidth]{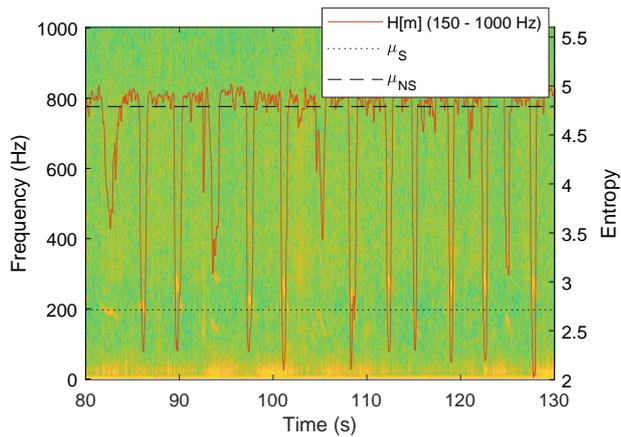}
	\caption{The resulting class means from the K-means algorithm on median filtered entropy with $M=355$.}
	\label{fig:kmeans1}
\end{figure}

An example of the resulting class means applied to the entropy measure is shown for the STFT in figure \ref{fig:kmeans1}.

\section{Simulation}

\subsection{Purpose}
Measuring classification accuracy on a per-sample basis is difficult to do with practical data, since the exact start and endpoints of a signal cannot be known exactly. Furthermore, the instantaneous SNR of a whale vocalization must be estimated, and does not remain constant.

As such, it is preferable to corrupt known test signals with noise for sample-accurate measurements. An exact SNR also allows for detailed investigation on the effect of SNR on the algorithm, which grants insight into designing a detector for practical data with estimated parameters.

Simulation is used to determine the best TF-decomposition for the practical data with estimated instaneous SNRs and provide insight on how to choose the median filter window length.

\subsection{Setup}
A pulsed FM sweep ($x[n]$) from 150 - 800 Hz is mixed with a sample of noise ($w[n]$) collected in the same time frame as the humpback songs from False Bay. The SNR is calculated only with respect to relevant indices used for entropy calculation, which ranges from 150 - 1000 Hz. \changes{As such, the SNR described in this section is higher than what would be observed practically.} This eliminates anchor noise, which contains most of the signal power, from the SNR calculation. The SNR is calculated using the segments where $x[n]$ is present.
 
\begin{figure}
 	\centering
 	\includegraphics[width=0.6\textwidth]{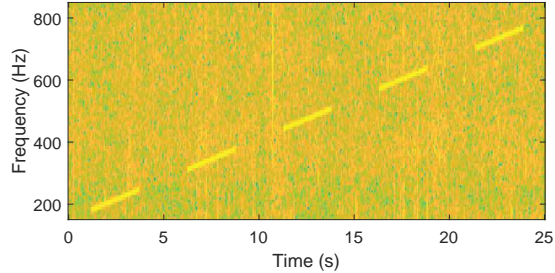}
 	\caption{Spectrogram of the test signal (0 dB SNR).}
 	\label{fig:testsigstft}
 \end{figure}

For this simulation, $x[n]$ is non-zero in 10\% of noise samples, pulsed evenly over a 25 second duration, sampled at 2 kHz. \changes{The STFT of the test signal is shown in figure \ref{fig:testsigstft}.}
 
All STFT's are performed with a Hamming window of length 256 with an overlap of 255. No zero-padding is applied for the FFTs. Figure \ref{fig:testsigstft} shows the STFT of the test signal mixed with the ocean noise with a SNR of 0 dB.
 
 All CWTs are L1-normalized and are performed with a Morse wavelet with parameters $\gamma = 50$, $\beta = 2000$, over the same frequency range as the STFT (150 - 1000 Hz). 40 filters per octave are used, corresponding to a total of 110 band-pass filters, which is a comparable number of frequency bins used by the STFT entropy calculation (109 bins). The high time-bandwidth product ensures that the filters are narrow enough to reduce contamination between frequency bins. The hyper parameter $g$ is set such that $p = 0.99$. 
 
\subsection{Measuring Soft-Output Accuracy}

Given an observed signal $y[n] = w[n] + x[n]$ of length $N$, where $w[n]$ is noise and $x[n]$ is a sporadically appearing signal. The presence of $x[n]$ is encoded by the set $S_x = \{ {n_1, n_2, ..., n_L} \}$, where $n_i \in \{0, 1, ..., N -1\}$ is the indices at which the signal is present. The cardinality (size) of the set is denoted as $|S_x|$.

The soft-output classification function $c_s(H[m])$ is used to determine the soft true positive (STPR) and true negative (STNR) discovery rates:
\begin{equation*}
	\text{STPR} = \frac{1}{|S_x|} \sum_{n \in S_x} c_s(H[m])
\end{equation*}
\begin{equation*}
	\text{STNR} = \frac{1}{N - |S_x|} \sum_{n \notin S_x} 1 - c_s(H[m])
\end{equation*}

This is equivalent to calculating the conventional TPR and TNR discovery rates if $c_h(H[m])$ is used instead of $c_s(H[m])$.
 
 \subsection{Results}
 \label{sec:fmsweep} 

\begin{figure}[h!]
 	\centering
 	\includegraphics[width=0.6\textwidth]{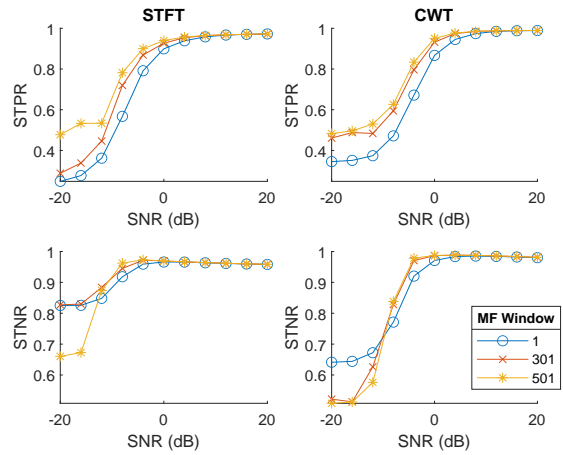}
 	\caption{STFT and CWT soft discovery rates over a range of SNR's and median filter window lengths. The left-hand side of the figure shows the STPR and STNR of the STFT respectively. The CWT soft discovery rates are on the right-hand side.}
 	\label{fig:res_snr_mf}
 \end{figure}

Figure \ref{fig:res_snr_mf} shows the STPR and STNR measures for various SNRs and median filter window lengths.
 
\begin{figure}[h!]
	\centering
	\includegraphics[width=0.6\textwidth]{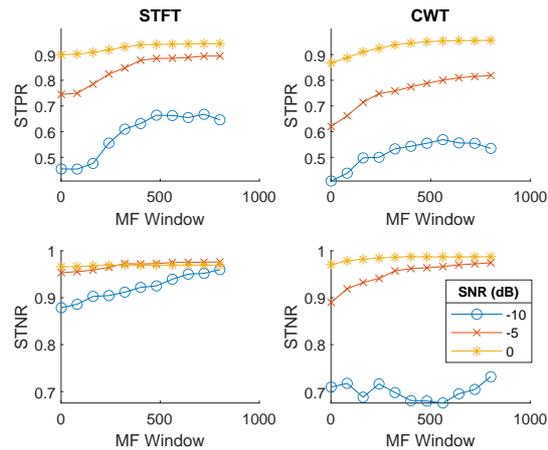}
	\caption{STFT and CWT soft discovery rates over a range median filter window lengths for low SNRs. The left and right sides of the figure show the results for the STFT and CWT respectively.}
	\label{fig:res_mf_snr}
\end{figure}

Figure \ref{fig:res_mf_snr} shows the STPR and STNR measures for median filter window lengths for SNRs of -10, -5 and 0 dB. As shown in tables \ref{tab:stpr} and \ref{tab:stnr}, the discovery rates for SNRs above 0 dB are greater than 90\%, with median filtering only showing minor improvements of the measures. Median filtering is therefore critical for lower power signals, which may include the fringes of a whale call waveform, due to ramping in volume.

Median filtering increases the STPR for low SNRs. Below -10 dB, it has a negative effect on the STNR. This is acceptable, since most practical signals are expected to operate above -10 dB SNR.

\begin{figure}[h!]
	\centering
	\includegraphics[width=0.6\textwidth]{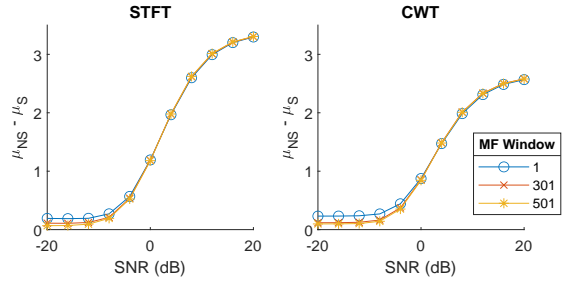}
	\caption{STFT and CWT difference in class means ($\mu_{NS} - \mu_S$) over a range of SNRs and median filter window lengths.}
	\label{fig:res_classmeans}
\end{figure}

Figure \ref{fig:res_classmeans} shows how the class mean separation is affected by SNR and median filter window length for the STFT and L1-normalized CWT. Median filtering has no significant influence on class separation above 0 dB SNR, but negatively impacts the separation below 0 dB SNR. The STFT exhibits better class mean separation compared to the CWT. Despite this, tables \ref{tab:stpr} and \ref{tab:stnr} show superior CWT performance, due to the better time localization properties of the CWT.

\begin{table}[h!]
	\caption{Comparing the STPR of the STFT and CWT at varying SNR's and MF Window Lengths\label{tab:stpr}}
	\centering
\begin{tabular}{cclllll}
\hline
	\multirow{3}{*}{} & \multicolumn{1}{c}{\multirow{3}{*}{M}} & \multicolumn{5}{c}{\textbf{Soft True Positive Rate (\%)}}                   \\
	& \multicolumn{1}{c}{} & \multicolumn{5}{c}{SNR}                  \\ \cline{3-7} 
	& \multicolumn{1}{c}{} & -10  & -5           & 0    & 5   & 10   \\ \hline
\multirow{3}{*}{STFT} & 1                    & 45.5          & 74.5          & 89.9          & 94.5          & 96.3          \\
& 301                  & 59.3          & 84.3          & 92.7          & 95.7          & 96.6          \\
& 501                  & \textbf{66.4} & \textbf{88.4} & 94.0          & 96.0          & 96.7          \\ \hline
\multirow{3}{*}{CWT}  & 1                    & 40.8          & 62.0          & 86.7          & 95.5          & 98.1          \\
& 301                  & 52.4          & 75.0          & 93.2          & 97.8          & 98.7          \\
& 501                  & 56.9          & 79.1          & \textbf{95.1} & \textbf{98.0} & \textbf{98.7} \\ \hline
\end{tabular}
\end{table}

\begin{table}[h!]
	\caption{Comparing the STNR of the STFT and CWT at varying SNR's and MF Window Lengths\label{tab:stnr}}
	\centering
	\begin{tabular}{cclllll}
		\hline
		\multirow{3}{*}{} & \multicolumn{1}{c}{\multirow{3}{*}{M}} & \multicolumn{5}{c}{\textbf{Soft True Negative Rate (\%)}}                   \\
		& \multicolumn{1}{c}{} & \multicolumn{5}{c}{SNR}                           \\ \cline{3-7} 
	& \multicolumn{1}{c}{} & -10  & -5           & 0    & 5   & 10   \\ \hline
\multirow{3}{*}{STFT} & 1                    & 87.9          & 95.3          & 96.6          & 96.5          & 96.2          \\
& 301                  & 91.2          & 97.2          & 96.9          & 96.6          & 96.3          \\
& 501                  & \textbf{92.9} & \textbf{97.4} & 96.9          & 96.6          & 96.3          \\ \hline
\multirow{3}{*}{CWT}  & 1                    & 71.0          & 89.0          & 97.1          & 98.5          & 98.5          \\
& 301                  & 70.8          & 95.6          & 98.7          & 98.8          & 98.7          \\
& 501                  & 66.8          & 96.4          & \textbf{98.7} & \textbf{98.8} & \textbf{98.7} \\ \hline
	\end{tabular}
\end{table}

The CWT exhibits poorer STNR and STPR performance for SNRs below 0 dB. Utilizing the equivalent sinusoidal power from equation (\ref{eqn:cwtps}) unevenly weights some of the frequency bins according to filter bandwidth, which lowers the CWT's performance for low SNRs, since the noise power becomes significant compared to the signal power. This makes the L1-normalized CWT ineffective for low SNRs. The reduction in class mean separation compared to the STFT in figure \ref{fig:res_classmeans} may also be explained by this phenomenon. 

For SNRs below 0 dB, it is recommended to use the STFT with a large MF window ($M \in [500, 1000]$). Note that this will reduce the capability of detecting shorter signals. For SNRs above 0 dB, the CWT performs best, with a recommended window size greater than 300. 

Median filtering has little to no significant effect above 10 dB SNR. If no median filtering is to be used ($M=1$), the STFT is recommended for SNRs below 5 dB and the CWT for SNRs above 5 dB. For a change in sampling frequency, the MF window lengths may be adjusted accordingly.

\subsection{A Note on L2 Normalization}

\begin{figure}[h!]
	\centering
	\includegraphics[width=0.6\textwidth]{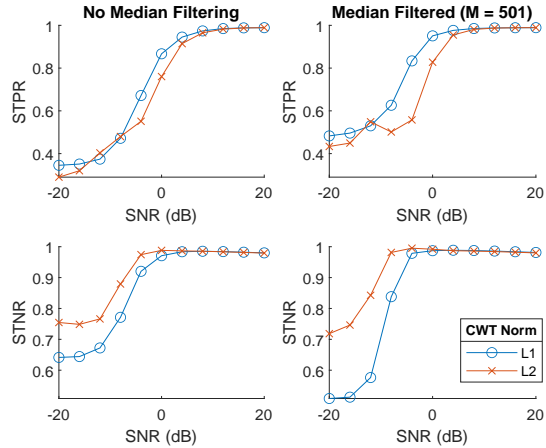}
	\caption{Comparing L1 and L2 CWT normalization for spectral entropy calculation. The left side of the figure shows the results without MF ($M=1$). The right side of the figure shows with MF ($M=501$). Note the large differences between L1 and L2 normalization for STNR and STPR.}
	\label{fig:res_snr_mf_L2}
\end{figure}

Figure \ref{fig:res_snr_mf_L2} compares the performance of the L1 and L2-normalized CWTs. The L2-CWT shows better STNR performance than the L1-CWT, since the power scaling effect is not present. However, the L2-CWT does not perform as well on STPR, due to the suppression of sinusoidal power at higher frequencies. 

The the STFT outperforms the L2-CWT on all accounts, making the L1-CWT the superior option, since it performs better than the STFT for higher SNRs.

\section{Application to the False Bay Data}

From simulation, it is determined that a larger median filter length and the L1-normalized CWT performs best for SNRs above 0 dB. As such, a median filter window lengths of 201 and 501 are tested for the proposed soft-classification technique, and compared to the baseline STFT-based SE (Hamming window of 256 samples; 255 sample overlap) used in \cite{entropyJASA}. No median filter is also tested on the CWT so that the STFT and CWT may be directly compared without enhancements. The detectors operate in the 130-1000 Hz frequency region, with the CWT having 40 filters per octave, $\gamma = 50$ and $\beta = 2000$.

The manual annotations are considered as the ground truth. However, it should be noted that these annotations may suffer from slight inaccuracies on a per-sample basis, which would reflect as a constant offset in the results, thus not contaminating this experiment significantly.

 \begin{figure}[h!]
	\centering
	\includegraphics[width=1\textwidth]{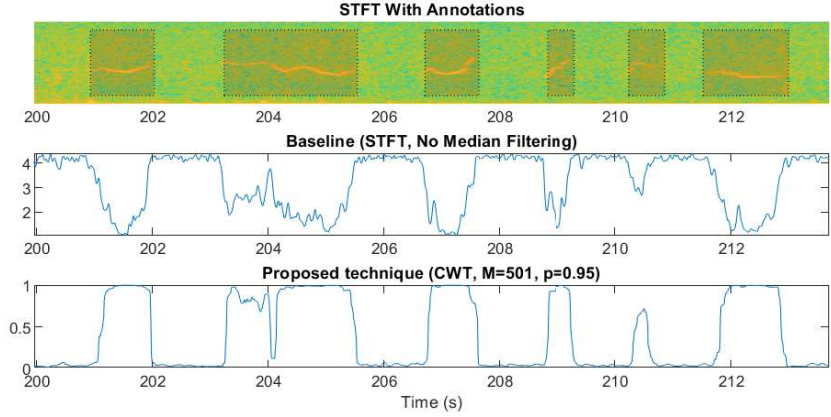}
	\caption{A snippet showing the measures produced by the entropy and the proposed soft-output signal detector.}
	\label{fig:res_realdata_snippet1}
\end{figure}

A receiver operating characteristic (ROC) curve is used to show the performance of the proposed technique, as is done in \cite{entropyJASA}. This curve plots the true positive rate (TPR) and the false positive rate (FPR) on the same axes, determined by sweeping over a detection threshold parameter. For the baseline SE, this threshold is swept linearly between the maximum and minimum entropy values. The proposed technique sweeps the pseudo-probability threshold from 0 to 1. The sigmoid gain is set such that $p=0.95$. Figure \ref{fig:res_realdata_snippet1} shows the output the baseline SE and the proposed detector on a snippet of the recorded humpback song.

The ROC curve illustrates the trade-off between probability of detection (TPR) and probability of false alarms (FPR) for a varying detection threshold. Curves which are closer to the upper left corner of the ROC plot indicate greater detection capability, since it reduces the risk of a false detection while remaining accurate (true detections). Accuracy is measured on a per-sample basis, since accurate start and endpoints are important for many applications.

The TPR and FPR measures are calculated using:
\begin{equation*}
	\text{TPR} = \frac{\text{\# True Positives}}{\text{\# Signal Samples}};
\end{equation*}
\begin{equation*}
	\text{FPR} = \frac{\text{\# False Positives}}{\text{\# No-Signal Samples}}.
\end{equation*}

\begin{figure}[h!]
	\centering
	\includegraphics[width=0.5\textwidth]{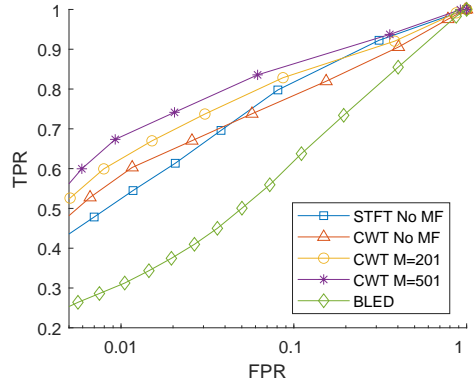}
	\caption{The ROC curves comparing the proposed technique and the baseline STFT SE. Median filter window lengths of 1, 201 and 501 are shown. The sigmoid gain is set such that $p=0.95$. FPR is plotted on a log scale, which better illustrates the TPR differences for small FPR values.}
	\label{fig:res_realdata_roc}
\end{figure}

\begin{figure}[h!]
	\centering
	\includegraphics[width=0.55\textwidth]{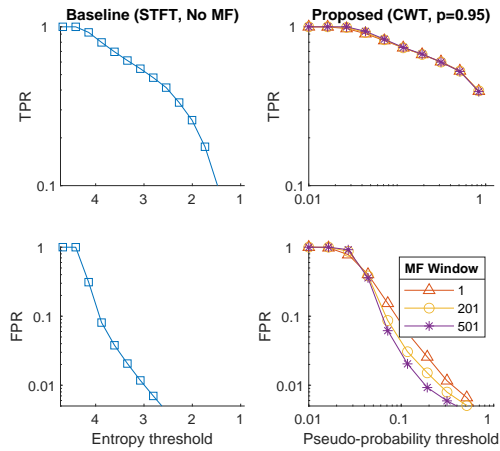}
	\caption{The thresholds used to produce the ROC curve and the TPR/FPR of the baseline SE and the proposed technique. Median filter window length of 1, 201 and 501 are shown on the right. The baseline entropy is shown on the left. }
	\label{fig:res_realdata_thresh}
\end{figure}

The ROC curve shown in figure \ref{fig:res_realdata_roc} is constructed from the two recording segments containing the humpback songs. The thresholds which produce the ROC curve is shown in figure \ref{fig:res_realdata_thresh}. \changes{For reference, the ROC curve of BLED is also shown, since it is widely used as a comparison in relevant literature \cite{internalsignaldetection, entropyJASA}. BLED is configured to calculate the energy in the 130-1000 Hz region via summation of the STFT produced for the baseline SE detector.}

From figure \ref{fig:res_realdata_roc}, it is clear that the proposed technique which uses a median filtering window of 501 samples and the CWT as a TF decomposition outperforms the conventional STFT-based SE. The proposed method shows an increase of around 10\% in sensitivity (TPR) for FPRs  below 3\%, and remains a better detector overall. \changes{As shown previously \cite{entropyJASA}, SE greatly outperforms a standard BLED detector.}

In practice, the detector should be tuned such that the FPR is below 5\%. From figure \ref{fig:res_realdata_thresh}, this corresponds to a threshold of $c_s(H[m]) > 0.1$ for $p=0.95$. This threshold will vary depending on the choice of $p$. Also note that the CWT SE without median filtering ($M=1$) outperforms the STFT in the $\text{FPR} < 5\%$ region, indicating that it is the best TF-decomposition for similar practical applications.

\section{Discussion}
\changes{The proposed detector based on the CWT with median filtering outperforms the traditional SE detector as described in \cite{entropyJASA} for practical applications, verified by real and simulated experiments. The comparison method is done on a per-sample basis, which also indicates increased time localization capabilities of the proposed detector. This is to be expected, since the CWT improves time localization for higher frequencies as a trade-off for filter bandwidth, as opposed to the STFT which has a fixed time-localization as a function of window length \cite{waveletbook}. Median filtering significantly decreases the FPR by suppressing fluctuations in the SE measure. Overall, the proposed detector indicates an approximate 10\% increase of TPR for FPRs below 3\%. As shown by \cite{entropyJASA}, we have again demonstrated that SE detectors outperform BLED on all accounts.}

\changes{It should be noted that the results shown by the ROC curve from figure \ref{fig:res_realdata_roc} are lower compared to Erbe and King's results \cite{entropyJASA}. Our paper's method of measuring performance differs from much of classification literature, since it constrains the window overlap to produce a TF decomposition for every sample, whereas classification accuracy is usually measured on a per-window basis \cite{hmm1, hmm2, hmm3, MLcnn, detectionGLRT} with less overlap. For simple classification problems, accurate audio segmentation and time localization is not necessarily a priority. In \cite{entropyJASA}, the window overlap is not maximal, thus detections are limited to the time localization accuracy determined by the overlap length. The obtained results are more similar to \cite{entropysubband}, which also measure accuracy on a per-sample basis.}

\changes{This paper introduces the k-mean clustering algorithm as a method to provide a pseudo-probabilistic measure of signal presence. This idea may be adapted to any binary signal detector, such as BLED \cite{internalsignaldetection}, to assist in providing a more intepretable measure of signal presence which does not depend on the magnitude of test statistic produced by the detector. The pseudo-probabilistic measure yields a threshold setting which may be more intuitive to non-experts, while also providing a value to be used to create adaptive algorithms which deal with sporadically appearing signals. Possible fields of application include denoising \cite{denoising, denoiseSSN}, instantaneous frequency estimation \cite{frequencyestimation} and classification (for example, as an additional input to a deep neural network) \cite{MLcnn}. Additionally, the CWT may offer better computational efficiency compared to the STFT if per-sample signal segmentation accuracy is required \cite{fcwt}, which could be relevant in preparing datasets for supervised learning.}

\changes{The addition of the k-means algorithm presents another advantage: utilizing the class-mean separation ($\mu_{NS} - \mu_S$) as metric to quickly determine whether there are points of interest within an audio segment. Furthermore, the k-means algorithm may be restricted to converge within a set number of iterations. Whether the algorithm converged indicates if there are points of interest within an audio segment. By segmenting the long recording and testing for appropriate class mean separation and k-means algorithm convergence, the proposed technique provides a fast way of discarding large portions of noise in long audio recordings. This method may be robust to changes in noise conditions, as long as the changes in noise are not experienced in a given segment of a long audio recording. The use of such a system remains to be explored.}

\changes{Automatic threshold setting is a possible extension of this study. This may be achieved by fitting a 2-component GMM to the SE distribution, with k-means clustering acting as the seed values of the fitting procedure, similar to HMM literature \cite{hmm1, hmm2, hmm3}. If a good approximation of the SE distribution is achieved, an appropriate threshold can be automatically determined for a fixed FPR specified by the user \cite{vantrees}.}

\changes{A SE-based detector may not be appropriate for detection in all whale species, since it is restricted to detecting vocalizations with periodic oscillations. This implies that some broadband impulsive noises, such as very short clicks, may not be detected. However, the detector will be insensitive to impulse noise disturbances, such as snapping shrimp \cite{denoiseSSN, internalsignaldetection}.}

\section{Conclusion}

\changes{The performance of whale vocalization detection in the presence of noise based on spectral entropy has been evaluated in this paper. Through simulation and comparison with manual annotations, the effectiveness of the CWT as a TF-decomposition for SE calculation is demonstrated. The use of median filtering also indicates significant improvement of the conventional STFT-based SE detector. Through simulation, this paper shows that the L1-normalized CWT outperforms the STFT and L2-normalized CWT for practical SNRs.}

\changes{The performance of the proposed detector shows an increase of approximately 5-10\% in accuracy (TPR), for false alarm rates (FPR) below 5\%, which is ideally what the detector should be tuned to. Furthermore, we propose a method of converting the SE entropy measure to a pseudo-probability using k-means clustering and automatic scaling, which is both more interpretable to non-experts while providing a foundation for developing adaptive algorithms for sporadically appearing cetacean vocalizations, which must take signal presence into account.}

\section{Acknowledgements}
\changes{This research is funded by the National Research Foundation (NRF) of South Africa (grant numbers: MND210609609887; 129224). The findings of this study and the opinions reflected therein are those of the authors and do not necessarily represent the NRF.}

\appendix
\section{K-means Implementation}
\label{app:kmeans}
For a known signal and noise profile, such as white noise, it is possible to estimate the expected values of the spectral entropy measure $H[m]$ given signal/no-signal conditions. However, in a real-world setting, the noise profile cannot be assumed to be a stationary process such as Gaussian white noise (GWN) or colored white noise (CWN). The variability of AMFM signals also further complicates this estimation. Setting appropriate thresholds on the entropy measure have to be done either manually, or via an automatic process. 

K-means is used to estimate these expected values in real-world conditions. A spectral entropy measurement $H[m]$ may belong to 2 classes: no signal ($C_{NS}$); signal ($C_S$). Note that it is required to have signal segments from both classes present in $H[m]$, otherwise this process will fail. It is assumed that the data satisfies the condition $\mathbb{E}\{H[m] | C_{S}\} < \mathbb{E}\{H[m] | C_{NS}\}$, where $\mathbb{E}$ denotes the expected value, implying that signal presence will always lower the entropy measure.

Class-membership is assigned using a hard class-assignment function:

\begin{equation*}
	w[m] = \begin{cases}
		1, & H[m] \in C_S\\
		0, & H[m] \in C_{NS}
	\end{cases}.
\end{equation*}

Given the class means $\mu_S$, $\mu_{NS}$ and the entropy measure $H[m]$ of length $N$, the assignment function assigns the class with the closest mean:

\begin{equation} 
\label{eqn:classassign}
	w[m] = \begin{cases}
		1, & (H[m] - \mu_S)^2 \leq (H[m] - \mu_{NS})^2 \\
		0, & \text{otherwise}
	\end{cases}.
\end{equation}

The K-means algorithm for this application is shown below:
\begin{enumerate}
	\item Assign initial means:
	\begin{equation*}
		\mu_S = \min\limits_{m}\left(H[m]\right);
	\end{equation*}
	\begin{equation*}
		\mu_{NS} = \max\limits_{m}\left(H[m]\right).
	\end{equation*}
	\item Perform class assignment using equation (\ref{eqn:classassign}).
	\item Update the means according to:
		\begin{equation*}
			\mu_S = \frac{\sum_{m=0}^{N-1} w[m] H[m]}{\sum_{m=0}^{N-1} w[m] };
		\end{equation*}
		\begin{equation*}
			\mu_{NS} = \frac{\sum_{m=0}^{N-1} (1-w[m]) H[m]}{\sum_{m=0}^{N-1} 1 - w[m] }.
		\end{equation*}
	\item Repeat steps 2 and 3 until the maximum change in means converges to a specified value $\epsilon$. Terminate if it exceeds a predefined number of iterations.
\end{enumerate}

Step 1 guarantees the means converge to their correct classes, and do not swap around in the process. It assumes that the maximum and minimum are not too extreme in that only a single point is associated with the initialization, and that signal/no-signal segments exist. The difference in class means may be used to determine whether a signal is present in the audio clip at all, since the class means will converge to similar values if no signal segments exist.





\bibliographystyle{elsarticle-num-names} 
\bibliography{cas-refs}





\end{document}